\documentclass[amssymb, amsmath, 11pt, showpacs, notitlepage]{revtex4}
\selectfont

\begin{document}

\title{Chiral Dirac--Born--Infeld solitons with SDiff symmetry}

\author{{\L}ukasz Bednarski}
\affiliation{Institute of Physics, Jagiellonian University, Reymonta 4, 
	Krak\'ow, Poland}


\begin{abstract}
	We propose a new DBI extension of a Skyrme type model which allows for
	BPS topological solitons with arbitrary value of the baryon charge. 
	The model is built out of the baryon density squared and in the limit
	of small fields tends to the BPS Skyrme model. We consider some 
	generalizations to higher dimensions and other K-deformed actions.  
\end{abstract}


\maketitle

\section{Introduction}

Constructing the low energy effective theory of Quantum Chromodynamics is of great interest as 
QCD can be solved perturbatively only at high energies. Although the correct form of the theory 
is still unknown it can be shown that the proper low energy degrees of freedom are encoded into 
mesonic fields, at least in the large $N_c$ limit \cite{Nc}. Then, baryons should appear as some collective 
excitations in mesonic `fluid'. One of the most successful realizations of this concept 
is the Skyrme model~\cite{skyrme}:
\[
	\mathcal L_{\rm Sk} = - \frac{f_\pi^2}{4} {\rm Tr\,} \left( L_\mu L^\mu \right)
		- \frac{1}{32e^2} {\rm Tr\,} \left( \left[ L_\mu ,
			L_\nu  \right]^2 \right) - \frac{m^2}{2} {\rm Tr\,}(1-U).
\]
where $U$ is a $SU(2)$ valued matrix meson field and $f_\pi$, $e$ are coupling constants. 
Further, $L_\mu = U^\dagger \partial_\mu U$. The standard $\sigma$--model alone (the first 
term of the Lagrangian) does not pass
Derrick's argument and therefore the second term, so-called Skyrme term is needed. The last 
part i.e., the Skyrme potential is optional from the stability point of view but rather 
necessary for phenomenological applications. For finite energy solutions one has to 
impose the following boundary condition:
$U(\infty) = 1\!\!1$, which allows for one--point compactification of base space: 
$\mathbb R^3 \cup \{\infty\} \cong \mathbb S^3$. Hence, for a fixed time $t$, 
$U$ is a map $\mathbb S^3 \to \mathbb S^3$ (target space $SU(2)$ is diffeomorphic
to $3$--sphere) and can be indexed by a topological 
number $B \in \pi_3 (\mathbb S^3) \cong \mathbb Z$, which is further identified 
with the baryon number \cite{witten}. This number is also the degree of the map $U$ and can be 
written explicitly as:
\[
	B = -\frac{1}{24 \pi^2} \int {\rm d}^3 x \, \varepsilon^{ijk} {\rm Tr\,}
		\left( L_i L_j L_k \right).
\]
In fact, skyrmions with higher baryon charge have been found in the massless as well 
as in the massive case~\cite{skyrmions}. Asymptotically, when $B\rightarrow \infty$,
one finds a skyrme crystal. 

It is worth to underline that the stabilizing static chiral solitons by the Skyrme term is not 
unique. There were proposed other chiral models with finite energy solutions. Usually, 
one introduces more fields (e.g. coupling with $\omega_\mu$ meson \cite{jackson}) or changes 
the form of the Lagrangian (e.g. Deser--Duff--Isham model~\cite{ddi}). Perhaps the most popular 
modification is of the Dirac--Born--Infeld (DBI) type~\cite{pavl1, nastase, ram}:
\[
	\mathcal{L}_{\rm DBI \; Sk} = -f_\pi^2  \; \beta^2 
		\left(1-\sqrt{1+\frac{1}{2\beta^2} \mbox{Tr}  L_\mu L^\mu}  \right) \;,
\]
where $f_\pi$ fixes the energy scale and  $\beta$ is a parameter of the model (we have chosen a length scale $l$ such that the coordinates $\vec{x}$ are dimensionless). For some generalizations see \cite{marleau, ram1}.
However, on the contrary to the standard skyrmions, little is known about DBI chiral solitons. 
It has been shown that the usual hedgehog ansatz works for $|B|=1$ solution
leading to a spherically symmetric, power-like localized DBI skyrmion.
Unfortunately, application of this ansatz to higher charge solutions gives
configurations (which are not critical points of the full energy functional) for
which energy grows as $|B|^3$. However, this result does not contradict the
existence of higher DBI skyrmions. It simply shows that the hedgehog ansatz is
not suitable for higher charge solutions. Therefore it is not known whether
higher DBI skyrmions do exist at all. What can be shown is that there is
 a linear topological bound:
\begin{equation}
	E_{\rm DBI \; Sk} \geq \frac{C f_\pi^2 }{\beta} |B|,
\end{equation}
where $C$ is a constant (see Appdendix).  

This is the first aim of the present paper to establish the existence of some DBI chiral solitons 
for an arbitrary value of the topological charge in an analytical way. In order to do that we 
modify the DBI Lagrangian in a way which enhances symmetries of the model i.e., the sigma model 
part $L_\mu^2$ will be replaced by the topological current squared. Concretely, we will 
focus on the following chiral model:
\begin{equation}
	\mathcal L_{\rm DBI} = - \beta^2 \left( 1 - \sqrt{1 - \frac{1}{2 \beta^2} B_\mu B^\mu}
	 	\right) - \mu^2 V \;,
\label{eq:dbi-lag}
\end{equation}
where 
\begin{equation}
B^\mu = -\frac{1}{24}\epsilon^{\mu \nu \rho \sigma} \mbox{Tr} \; L_\nu L_\rho L_\sigma
\end{equation}
is the baryon current and $V=V(U,U^\dagger)$ is a potential multiplied by a constant $\mu^2$. Then, in other words, it is a 
DBI generalization of the recently proposed  \emph{BPS Skyrme model}~\cite{bps-skyrme}, which 
seems to provide the correct starting point for the low energy effective action of QCD (it 
leads to physical binding energies \cite{energy-bps} and Roper spectra \cite{rot-vib-bps}; see also \cite{marl}).
In fact, in the limit $\beta^2 \to \infty$ one finds the BPS Skyrme Lagrangian:
\begin{align}
	\mathcal L &= -\pi^4 \lambda^2 B_\mu B^\mu - \mu^2 V.
			\label{eq:bps-skyrm-lag} 
\end{align}
Hence, the present paper will also help to understand what happens  with BPS skyrmions 
if the action is modified to the DBI type. This issue can be analyzed from a wider 
perspective as we want to start with the most general chiral model built out of the baryon density squared:
\begin{equation}
 \mathcal{L}= \mathcal{L}(B^2_\mu, U, U^\dagger) \;.
\end{equation}
Such a broad class of models with rather nonstandard kinetic terms, usually called as K field theories, 
has been recently studied especially in the context of cosmology. They offer a solution to problems 
concerning inflation (K inflation \cite{inflation}) or late time acceleration (K essence \cite{essence}) 
and allow for new topological defects \cite{defects} which relevance for application to the 
structure formation in the early Universe is investigated. 
\section{Generalized models with SDiff symmetry}
\subsection{Definition and symmetries}
Let us make our investigation even more general and begin with an arbitrary $(d+1)$ dimensional 
space-time and a set of $d$ scalar fields $\phi^a$, $a = 1, 2, 
\ldots, d$. We assume that target space $\mathcal{M}$ is Riemannian space and line element
is given by: ${\rm d} s^2 = g_{ab} \, {\rm d}\phi^a \, {\rm d}\phi^b$. We then
define topological current $B^\mu$ as a pullback of volume form on target 
space \cite{bps}:
\begin{equation}
	B^\mu = M(\phi^a) \; \varepsilon^{\mu \mu_1 \ldots \mu_d}
			\partial_{\mu_1} \phi^1 \ldots \partial_{\mu_d} \phi^d \;, 
	\quad
	M(\phi^a) \equiv \sqrt{\det g_{ab}} \;.
\end{equation}
Then, the K--generalized BPS models are defined by the following Lagrangian:
\begin{equation}
	\mathcal{L} = \mathcal{L}(B_\mu^2, \phi^a)\;. \label{kbps}
\end{equation} 
Obviously, symmetries of the model strongly depend on the particular form of the $\mathcal{L}$ function. 
However, as topological current squared is obviously invariant under the infinite group of 
volume--preserving diffeomorphisms on target space, such a K--BPS model may still possess 
infinitely many symmetries (which correspond to infinitely many Noether conserved charges). 
As an example, we are going to further analyze:
\begin{equation}
\mathcal{L}=F(B_\mu^2) - V(\phi^a) \label{kbps1}\;.
\end{equation} 
where $V$ is a potential. Then, invariance of $V$ under a
certain subgroup of ${\rm SDiff} (\mathcal{M})$ implies that the full model has infinitely many conservation laws. 

One can also notice that the energy functional of the static configurations is invariant 
under another infinitely large group which is volume preserving diffeomorphisms of 
base space. It means that the moduli space of (\ref{kbps}) is infinitely dimensional. 

\subsection{Static equations of motion}
In static case, the Lagrangian \eqref{kbps} simplifies considerably as
then there is only one non-zero component of topological current, namely the charge density $B_0$. 
When static lagrangian depends only on the $B_0$ and fields $\phi^a$, the 
second order Euler--Lagrange equations of motion can be integrated once and 
reduced to the first order differential equation. To see this, let us
consider more general form of a static energy density, namely:
\begin{equation}
	\mathcal{E} = F \left( A, \phi^a \right), \quad A = \varepsilon^{\mu_1 \mu_2
					\ldots \mu_d} \phi^1_{,\mu_1} \phi^2_{,\mu_2} \ldots
					\phi^d_{,\mu_d} \;.
\label{}
\end{equation}
The Euler--Lagrange equation is then:
\begin{equation}
	\partial_j \left( \frac{\partial F}{\partial A} \right) \cdot \frac{\partial
		A}{\partial \phi^b_{,j}} - \frac{\partial F}{\partial \phi^b} = 0 \;.
\label{eq:el-1}
\end{equation}
Applying the chain rule on the first term and simplifying the formula gives us:
\begin{equation}
	\partial_j A \cdot \frac{\partial^2 F}{\partial A^2} \cdot \frac{\partial
	A}{\partial \phi^b_{,j}} + A \frac{\partial^2 F}{\partial \phi^b \partial A}
	- \frac{\partial F}{\partial \phi^b} = 0 \;.
\label{eq:el-2}
\end{equation}
In the previous steps we have used the following helpful identities:
\[
	\partial_j \left( \frac{\partial A}{\partial
		\phi^b_{,j}} \right) = 0, \quad
	\frac{\partial A}{\partial \phi^a_{,j}} \cdot \frac{\partial
		\phi^b}{\partial x^j} = \delta^b_a A \;.
\]
We than assume existence of the BPS equation: $A = \pm W(\phi^a)$ and rewrite
\eqref{eq:el-2} ($A=W$):
\begin{equation}
	\frac{\partial^2 F}{\partial W^2} \cdot W \cdot \frac{\partial W}{\partial
		\phi^b} + W \cdot \frac{\partial^2 F}{\partial \phi^b \partial W} -
		\frac{F}{\phi^b} = 0 \;,
\label{}
\end{equation}
where $F \equiv F(W, \phi^a)$. After some basic manipulations we end with
\begin{equation}
	\frac{ {\rm d} }{ {\rm d} \phi^b } \left( \frac{\partial F}{\partial W}
		\cdot W	\right) - \frac{\partial F}{\partial W} \cdot \frac{\partial W}{\partial
		\phi^b} - \frac{\partial F}{\partial \phi^b} = 0 \;.
\label{}
\end{equation}
This can be integrated to:
\begin{equation}
	\frac{\partial F}{\partial W} \cdot W - F = 0 \;,
\label{eq:el-3}
\end{equation}
under the condition that both fields and their derivatives tend to $0$ at
spatial infinity. Then, using this formula one may find $W$ as a function of target space variables. 
Observe, that the existence of the BPS equation does not follow from any ansatz but is a general 
feature of the models built out of the topogical current squared. 

As a first example we consider the following family of K-BPS models of type (\ref{kbps1}):
\begin{equation}
\mathcal{E}= \left( B_0^2 \right) ^{\alpha} +\mu^2 V \label{kbps V}\;,
\end{equation}
where $\alpha$ is a parameter and $\mu$ a constant multiplying potential.  
Then, the BPS equation reads (here $B_0=MA$):
\begin{equation}
	\left(B^2_0\right)^{\alpha} (2\alpha -1)=\mu^2 V \quad \Rightarrow \quad 
	B_0= \left( \tfrac{\mu^2}{2\alpha -1}  V \right)^{\frac{1}{2\alpha}} \;.
\end{equation}
From the boundary conditions specified above we find a restriction of the parameter $\alpha > \frac{1}{2}$. 

For the more interesting case i.e., the DBI SDiff model (in $d$ dimensions):
\begin{equation}
	E = -\int {\rm d}^d x \, \mathcal{L}_{\rm static} =
		 \int {\rm d}^d x \left\{ \beta^2 \left( 1 - 
			\sqrt{1 - \frac{1}{2 \beta^2} B_0^2} \right)  + \mu^2 V  \right\} \;.
	\label{eq:stat-e}
\end{equation}
the BPS equation
is of the following form:
\begin{equation}
	\frac{1}{\sqrt{1 - \frac{1}{2 \beta^2} B_0^2}} = \frac{\mu^2 }{\beta^2} V+ 1 \;.
	\label{eq:dbi-bps}
\end{equation}
In the next sections this equation will be further analyzed. 
\subsection{Static Energy}
It is natural to expect that solutions of BPS equations will lead to linear relation between 
energy and topological charge. Here we prove that it is indeed the case. In order to do that one 
has to rewrite the static energy density $\mathcal{E}$ as the topological charge density 
multiplied by a function of target space coordinates. In general it is a rather complicated 
algebraical problem. Therefore, we show it for two previously defined types of models.

For the family \eqref{kbps V} we get:
\begin{align}
	E &= \int {\rm d}^d x \, \Bigl( \left( B_0^2\right)^\alpha + \mu^2 V \Bigr)
		= 2\alpha \left( \frac{2\alpha-1}{\mu^2} \right)^{\frac{1}{2\alpha}-1} 
			\int {\rm d}^d x \,  B_0 V^{1-\frac{1}{2\alpha}} \nonumber \\
	  & = 2 \alpha \left( \frac{2\alpha-1}{\mu^2} \right)^{\frac{1}{2\alpha}-1} |B| 
	  		\int_{\mathcal M'} {\rm d} \Omega^{(d)} V^{1-\frac{1}{2\alpha}}  \nonumber\\
	  &\equiv  2 \alpha \left(\frac{2\alpha-1}{\mu^2} \right)^{\frac{1}{2\alpha}-1} |B| 
	  	\left\langle V^{1-\frac{1}{2\alpha}}\right\rangle_{\mathcal M'} 
	  	\cdot {\rm vol} \mathcal M' \;.
		\label{eq:stat-e-kbps}
\end{align}
Restricting integration to a subspace $\mathcal M' \subset \mathcal M$ 
takes into account necessity of removing zeros of $B_0$ from integration region.
As it can be seen from \eqref{eq:dbi-bps} $B_0 = 0$ on a vacuum manifold, which
in our cases is one point.
Topological charge $B \in \mathbb Z$, which is equivalent to the winding number, has
been introduced as a consequence of the fact that the field $\vec \phi (\vec x)$
may cover $\mathcal M'$ region $B$ times while $\vec x$ covers base space once.
Furthermore, ${\rm d} \Omega^{(d)}$ is target space volume form.
\\
Analogously for the DBI SDiff model we find:
\begin{align}
	E &= \int {\rm d}^d x \, B_0 \cdot \frac{\beta^2 \left( 1 - 
			\sqrt{1 - \frac{1}{2\beta^2} B_0^2} \right) + \mu^2 V}{B_0} \nonumber\\
	&= |B| \sqrt{2} \beta  \int_{\mathcal M'} {\rm d}^d \phi \, M(\phi^a) 
			\sqrt{ \left( \frac{\mu^2 V}{\beta^2} \right)^2 + 2 \frac{\mu^2 V}{\beta^2}} \nonumber\\
	&= |B| \sqrt{2} \mu \int_{\mathcal M'} {\rm d} \Omega^{(d)}
			\sqrt{ \frac{\mu^2 V^2}{\beta^2} + 2 V} \nonumber\\
	&= |B| \sqrt{2} \mu \left\langle \sqrt{\frac{\mu^2 V^2}{\beta^2} + 2V}
		\right\rangle_{\mathcal M'} \cdot {\rm vol} \mathcal M'\;.
	\label{eq:stat-e-bps}
\end{align}
In both cases, the static energy is a linear function of the pertinent topological charge $B$. 
The proportionality constant is expressed as target space average value of a certain function 
of the potential, which encodes the nonlinearity of the derivative term. 
\\
From the BPS equation we may also observe that, in contrast to the usual Skyrme model \cite{krusch}, the topological density is always non-negative (for positive baryon charge). It again coincides with the property of the BPS Skyrme model. 
\section{DBI BPS models}
\subsection{DBI BPS baby Skyrmions}
As a first simple example of BPS models with DBI type action, we consider 
the DBI generalization of the BPS baby Skyrme model \cite{BPS baby} that is Lagrangian 
\eqref{eq:dbi-lag} in $d=2+1$ dimensions. Then, target space is given by
$\mathbb{S}^2$. Instead of considering fields on the 2--sphere (usual three
component unit vector field $\vec{\phi}$)
we use stereographic projection and introduce two complex fields $u$ and $\overline{u}$.
We than take topological current $B^\mu$ to be proportional to the pullback of 
the volume form on the target manifold ($\partial_\alpha u \equiv u_{,\alpha}$, etc.):
\begin{equation}
	B^\mu = - \frac{i}{2 \pi} \cdot \frac{\epsilon^{\mu \alpha \beta} u_{,\alpha}
		\overline{u}_{,\beta}}{(1 + |u|^2)^2} \;.
\label{}
\end{equation}
Both lagrangian and equation of motion can be written in more compact form if we introduce two auxiliary objects:
\begin{align}
	K_\mu &\equiv u_{,\alpha} \overline{u}^{,\alpha} u_{,\mu} - 
			u_{,\alpha} u^{,\alpha} \overline{u}_{,\mu}  \;, \\
	\mathcal K_\mu &\equiv \frac{K_\mu}{(1 + |u|^2)^2 \sqrt{1 - \frac{1}{8 \pi^2 \beta^2} 
		\frac{K_\nu \overline{u}^{,\nu} }{(1 + |u|^2)^4} 
		}} \;.
	\label{}
\end{align}
Then, e.o.m. can be written as:
\begin{equation}
	\partial_\mu \mathcal K^\mu - 8 \pi^2 \mu^2 (1 + |u|^2)^2 u V' = 0 \;.
	\label{}
\end{equation}
We assume here that the potential $V$ is a function of $|u|^2$, i.e. $V \equiv
V(|u|^2)$ and has only one vacuum value located at $u=0$. For example, one may consider family of generalized old baby potentials 
\begin{equation}
V = \left( \frac{|u|^2}{1+|u|^2} \right)^\alpha
\end{equation}
with $\alpha >0$. 
\\
In the static case we proceed with the ansatz: $u(r, \phi) = f(r) e^{in\phi}$, 
$f(r) \in \mathbb{R}$, where $n$ can 
be identified with topological charge. The proper boundary conditions corresponding 
to nontrivial topology are: $f(0) = +\infty$, $f(R) = 0$, where $R$ can 
be finite (compactons \cite{arodz}) or infinite.
To further simplify equations, we 
introduce new variables: 
\begin{equation}
	x = \tfrac{1}{2}r^2, \quad h = \frac{f^2}{1 + f^2} \;,
\end{equation}
where: $h(x) \in [0, 1]$, $h(0) = 1$ and $h(x_0) = 0$.\\
This is indeed helpful because then static e.o.m. can be written as:
\begin{equation}
	n^2 \frac{\rm d}{{\rm d}\,x} \frac{h_{,x}}{\sqrt{1 - \frac{n^2}{8\pi^2 \beta^2} h_{,x}^2 }}
		- 8\pi^2 \mu^2 \frac{{\rm d}\,V}{{\rm d}\,h} = 0 \;,
	\label{eq:h-eom}
\end{equation}
and can be further integrated to the first order differential equation:
\begin{equation}
	h_{,x} = - \frac{2 \sqrt{2} \pi \beta}{|n|} \sqrt{1 - \left( \frac{\mu^2}{\beta^2}
			 V + 1 \right)^{-2}} \;.
	\label{eq:hx-V}
\end{equation}
which is in a total agreement with the previously obtained BPS equation.
\\
Due to the fact that, in the near vacuum regime i.e., for small value of
topological charge density,  the DBI model tends to the (baby) BPS Skyrme 
model, we may easily understand how the type of solitons changes with the 
near vacuum asymptotic of the potential $V \approx h^\alpha$. There are 
three possibilities: (1) $\alpha \in (0,2)$ - compactons i.e., solitons 
approach the vacuum value ($h=0$) on a finite distance; (2) $\alpha=2$ - 
exponentially localized solitons; (3) $\alpha >2$ - power--like localized solitons.  
\\
As an exact example we shall consider the old baby potential that is the case when $\alpha=1$ $$ V= h.$$ Then, solutions are of a compacton type:
\begin{equation}
	h(x) = \begin{cases} \frac{\beta^2}{\mu^2} \left( \sqrt{1 + \frac{8\pi^2 \mu^4}
	{n^2 \beta^2} (x - x_0)^2} - 1\right) & \text{for} \; 0 \le x \le x_0 \;, \\
	0 & \text{for} \; x > x_0 \;,
	\end{cases}
	\label{eq:h-sol}
\end{equation}
where $x_0 = \frac{|n|}{2 \pi}\sqrt{ \frac{1}{2\beta^2} + \frac{1}{\mu^2}}$.

Explicit formula for the static energy in this configuration is:
\begin{align}
	E &= 2 \pi \int_0^{x_0} {\rm d}x \, \left\{ \beta^2 \left( 1 - 
			\sqrt{1 - \frac{n^2}{8 \pi^2 \beta^2} h_{,x}^2}
	\right) + \mu^2 h \right\} \nonumber \\
	&= |n| \cdot \pi \beta^2 \left[ \widetilde{x}_0 \sqrt{1 + v \, \widetilde{x}_0^2}
		 - \frac{1}{\sqrt{v}} \sinh^{-1} \left(\sqrt{v} \, \widetilde{x}_0 \right) \right] \;, 
	\label{eq:energy}
\end{align}
where we have introduced:
\[
	v = \frac{8 \pi^2 \mu^4}{\beta^2} \; {\rm and \;} \widetilde{x}_0 = \frac{x_0}{|n|}\;.
\]
This result is proportional to $|n|$ and coincides exactly with one computed 
using \eqref{eq:stat-e-bps}. As the static energy function is invariant under base 
space area preserving diffeomorphisms one can find other (energetically equivalent) 
solutions simply by applying these trasformations. 
\\
The solitonic solutions cease to exist when the potential term is absent. Indeed, then eq. \eqref{eq:hx-V}
reduces to: $h_{,x} = 0$. This cannot yield a continuous solution which
satisfies both boundary conditions: $h(0) = 1$ and $h(+\infty) = 0$. On the level of 
solutions (e.g. \eqref{eq:h-sol}) the limit $\mu \rightarrow 0$ 
leads to an almost flat (step-function) compacton which 
size goes to infinity. The corresponding energy
\eqref{eq:energy} goes to zero as:
\begin{equation}
	E = \frac{2|n|}{3}  \mu + \mathcal{O}(\mu^2)\;.
	\label{eq:E-mu}
\end{equation}
Hence, the static energy decreases linearly with $\mu$ and vanishes in the limit.

\subsection{DBI Skyrmions}
In 3+1 dimensional case we take target space to be $\mathbb{S}^3 \cong SU(2)$ 
and we choose the following parametrization of the $U \in SU(2)$ field:
\[
	U = e^{i \xi \vec n \cdot \vec \tau} = \cos \xi \, 1 \!\! 1 + i
			\sin \xi \, (\vec n \cdot \vec \tau) \;.
\]
$\xi$ is a real field, $\vec \tau$ is a vector of Pauli matrices and $\vec n \in
\mathbb{R}^3$ is a unit vector. As in a 2+1 dimensional case we relate $\vec n$ 
with two complex fields $u, \overline{u}$ by means of the stereographic projection.
Topological current thus reads:
\begin{equation}
	B^\mu = - \frac{i}{\pi^2} \cdot \frac{\sin^2 \xi}{(1 + |u|^2)^2}
			\varepsilon^{\mu \nu \rho \sigma} \xi_{,\nu} u_{,\rho}
			\overline{u}_{,\sigma} \;.
\end{equation}
We assume that the potential is of the form: 
\[
	V \equiv V \left( {\rm Tr\,} \left( U + U^\dagger \right) \right) = V (\xi) \;.
\]
In order to shorten our future notation, we introduce an abbreviation:
$P = 1 - \tfrac{1}{2\beta^2} B^\mu B_\mu$. We now proceed to equations of motion.
E.o.m. of the field $u$ (and its complex conjugate) simplifies vastly in its form after introduction of
the follwoing auxiliary quantity:
\begin{equation}
	\mathcal K^\mu \equiv \frac{1}{\sqrt{P} (1 + |u|^2)^2}  \cdot
		\frac{\partial}{\partial \overline{u}_{,\mu}} 
		\left( \varepsilon^{\alpha \nu \rho \sigma} \xi_{,\nu} u_{,\rho} \overline{u}_{,\sigma} \right)^2
		\;.
\end{equation}
It then can be written in a compact form as:
\begin{equation}
 	\partial_\mu \mathcal K^\mu = 0 \;.
 	\label{eq:uc-eom}
\end{equation}
When writing e.o.m. for the field $\xi$, it is also useful to introduce another 
auxiliary object, namely:
\begin{equation}
	\mathcal H^\mu \equiv \frac{\sin^2 \xi}{\sqrt{P}} \cdot
		\frac{\partial}{\partial \xi_{,\mu}} 
			\left( \varepsilon^{\alpha \nu \rho \sigma} \xi_{,\nu} u_{,\rho} \overline{u}_{,\sigma} \right)^2
			\;.
\end{equation} 
Hence, the second e.o.m. is:
\begin{equation}
	\frac{\sin^2 \xi}{(1 + |u|^2)^4} \partial_\mu \mathcal H^\mu 
		+ 4 \pi^4\mu^2 \, V_{,\xi} = 0 \;.	
	\label{eq:xi-eom}
\end{equation}
Static case simplifies greatly equations of motion. To solve them, we use the
axially symmetrical ansatz:
\begin{equation}
	\xi \equiv \xi (r), \quad u \equiv g(\theta) e^{in\phi} \;.
\end{equation}
$(r, \theta, \phi)$ are base space spherical coordinates. The first equation 
\eqref{eq:uc-eom} is then:
\begin{equation}
	g \cdot \partial_\theta \left(
		\frac{g g_{,\theta}}{\sqrt{P} (1+g^2)^2 \sin \theta}
	 \right) = 0 \;.
\end{equation}
Solution, which satisfies the correct boundary conditions $g(0) = 0$ and 
$g(\pi) = +\infty$, is simply: 
\begin{equation}
	g(\theta) = \tan \frac{\theta}{2} \;.
	\label{eq:g}
\end{equation}
We further use \eqref{eq:g} to simplify the second e.o.m. \eqref{eq:xi-eom}:
\begin{equation}
	\frac{n^2}{8 \pi^4} \cdot \frac{\sin^2 \xi}{r^2} \partial_r \left(
		\frac{\sin^2 \xi \cdot \xi_{,r}}{r^2 \sqrt{1 - \frac{n^2}{8 \pi^4 \beta^2}
		\cdot \frac{\sin^4 \xi \cdot \xi_{,r}^2}{r^4}}}
	\right) = \mu^2 V_{,\xi} \;,
	\label{eq:xi1}
\end{equation}
with the boundary condition: $\xi(0) = \pi$, $\xi(+\infty) = 0$.\\
We note that after substitution $z = \frac{2 \sqrt{2} \beta \pi^2}{|n|} r^3$ 
equation \eqref{eq:xi1} can be easily integrated:
\begin{equation}
	\frac{\beta^2}{\sqrt{1 - (\sin^2 \xi \cdot \xi_{,z})^2}} = \mu^2V + \beta^2 \;.
	\label{eq:xi2}
\end{equation}
When integrating, we made use of the condition: $V(\xi = 0) = 0$. Again, it is in perfect agreement with our BPS equation. 

As in the baby case, the type of solitons is governed by the asymptotic of the potential at the vacuum 
$V \approx \xi^\alpha$ and agrees with results for the usual BPS Skyrme model. 
Namely, (1) $\alpha \in (0,\tfrac{3}{2})$ we find compactons; 
(2) for $\alpha = \tfrac{3}{2}$ we have an exponentially localized solution; 
(3) for $\alpha > \tfrac{3}{2}$ we get solitons with power-like decay.  

As a particular exact example we consider the usual Skyrme potential:
\[
	V = \frac{1}{2} \mbox{Tr}\; (1-U)= (1-\cos \xi) \;.
\]
Then the formula for the profile function $\xi(z)$ (compacton) can be given in an implicit form:
\begin{equation}
\begin{cases}
	\frac{\sigma + \cos \xi}{2} \sqrt{(1 + \cos \xi)(1 + 2 \sigma - \cos \xi)}
		+ (1 - \sigma^2) \tan^{-1} \left( \frac{\sqrt{1 + \cos \xi}}{\sqrt{1 + 2
			\sigma - \cos \xi}} \right) = z, & {\rm for\;} z
			\le z_0 \\
	0, & {\rm for\;} z > z_0 \;,
\end{cases}
\label{eq:xi-sol3}
\end{equation}
where: $\sigma = \beta^2 / \mu^2$, $z_0 = \sqrt{\sigma} (1 + \sigma) + (1 -
\sigma^2) \tan^{-1} \tfrac{1}{\sqrt{\sigma}}$.\\
The asymptotics of this solution near the domain endpoints is as follows:
\begin{align*}
	\xi &\approx \frac{\sqrt{2(z_0 - z)}}{\sigma^{1/4}}, \quad\quad\quad z \to z_0^-, 
		\xi \to 0^+ \\
	\xi &\approx \pi - \sqrt[3]{6} \frac{\sqrt[6]{1 +
		\sigma}}{\sqrt[3]{2+\sigma}} \cdot z^{1/3}, 
		\quad z \to 0^+, \xi \to \pi^- \;.
\end{align*}
The apparent infinite jump of derivative at the boundary is smoothed in
expressions for the energy and baryon charge densities, in the sense that they
are analytic functions of $r$ variable.\\
The static energy is equal:
\[
	E = \frac{\sqrt{2} |n| \beta}{6 \pi \sigma} \left( \sqrt{\sigma} (12 +
\sigma(5+3\sigma)) + 3(1+\sigma) (\sigma^2 + \sigma - 4) \tan^{-1} \sqrt \sigma
\right) \;.
\]

Another simple example can be obtained when one takes:
\[
	V =  \tfrac{1}{2}(\xi - \cos \xi \sin \xi) \equiv \eta \;,
\]
which is usually referred as the BPS potential.
Then using \eqref{eq:xi2} we obtain:
\begin{equation}
	\eta_{,z} = - \sqrt{1 - \frac{1}{(\frac{\eta}{\sigma} + 1)^2}}
	\label{}
\end{equation}
which gives the following compact solution
\begin{equation}
\eta(z) = \begin{cases} 
	\sigma \left( \sqrt{1 + \left( \frac{z_0 - z}{\sigma} \right)^2} - 1 \right)  & {\rm for \;} z \le z_0, \\
	0 & {\rm for\;} z > z_0\;,
\end{cases}
\label{}
\end{equation}
where $z_0 = \frac{\sqrt{\pi}}{2} \sqrt{\pi + 4 \sigma}$, $\sigma = \beta^2 / \mu^2$
 and the static energy of this configuration:
\[
	E = \frac{\sqrt{2} \beta}{6 \pi} |n| \left( z_0 \sqrt{1 + \frac{z_0^2}{\sigma^2} }
		 - \sigma \sinh^{-1} \left(\frac{z_0}{\sigma}  \right) \right) \;.
\]\\
As in 2+1 dimensional case the potential term is crucial to obtain
nonzero solutions.
\section{Summary}
In this paper we have analyzed a new class of (3+1) dimensional Skyrme like models with the 
Dirac--Born--Infeld type Lagrangian. Specifically to overcome the technical problems affecting 
the usual Skyrme model (and its DBI modification) we built our Lagrangian using square of topological 
charge density. Therefore, one may consider the model as the DBI generalization of the 
BPS Skyrme model.  
Although the form of this DBI Lagrangian differs rather radically from 
the standard BPS Skyrme model some essential features remain unchanged. 
\\
Firstly, all topological solitons are solutions to a BPS equation which 
leads to a linear relation between the
energy and topological charge. Presented axially symmetric solitonic
 solutions are exact (up to an integral) which 
again is probably related to symmetries (see below) and the generalized 
integrability of the model \cite{gen}.
\\
Secondly, the model possesses exactly the same symmetries as the BPS 
Skyrme model. Specifically, there is an 
infinite symmetry group which is a subgroup of the volume preserving 
diffeomorphisms on target space 
$\mathbb{S}^3$. Further, the static energy functional is also
 invariant under base space volume preserving 
diffeomorphismsm of $\mathbb{R}^3$. Therefore, there are infinitely
 many solutions related to obtained ones by some VPD. 
\\
Thirdly, the potential term is unavoidable. DBI BPS solitons disappear
 as we approach the potential-less limit. 
This resembles the situation in the BPS Skyrme model and is in a striking 
contrast to the usual (3+1) dimensional DBI Skyrme model 
where potential is not obligatory for the existence of the solitons.     

Moreover, we extended some results to K-BPS models i.e., models which 
Lagrangians are any (reasonable) functions 
of the topological current squared with further generalization to 
any dimension. This open a way for searching 
twin models \cite{twin} for the usual BPS or DBI BPS Skyrme Lagrangians. 
\appendix
\section{}
We give the lower topological band on the static energy for the DBI model with the following lagrangian~\cite{pavl1,nastase}:
\[
	\mathcal{L} = - f_\pi^2 \beta^2 \left(1 - \sqrt{1 + \frac{1}{2 \beta^2} {\rm Tr} L_\mu L^\mu} \right) \;.
\]
We define the strain tensor~\cite{manton,harland,adam-wer}:
\[
	D_{jk} = -\frac{1}{2} {\rm Tr} L_j L_k \;,
\]
which is a positive and symmetric $3 \times 3$ matrix with non--negative eigenvalues $\lambda_1^2$, 
$\lambda_2^2$ and $\lambda_3^2$. The baryon number $B$ is given by:
\[
	B = \frac{1}{2\pi^2} \int {\rm d}^3 x \; \lambda_1 \lambda_2 \lambda_3 \;.
\]
We then write the formula for energy and expand the integrand into Taylor series: 
\begin{align}
	E &= f_\pi^2 \beta^2 \int {\rm d}^3 x \; \left(1 - \sqrt{1 - \frac{1}{\beta^2} \left(\lambda_1^2 + \lambda_2^2 +
		\lambda_3^2 \right)} \right) \nonumber \\
	& = f_\pi^2 \int {\rm d}^3 x \; \left( \frac{1}{2} \left(\lambda_1^2 + \lambda_2^2 +
		\lambda_3^2 \right) + \frac{1}{8 \beta^2} \left(\lambda_1^2 + \lambda_2^2 +
		\lambda_3^2 \right)^2 + \frac{1}{16 \beta^4} \left(\lambda_1^2 + \lambda_2^2 +
		\lambda_3^2 \right)^3 + ... \right) \;.
	\label{eq:app-1}
\end{align}
If we estimate this infinite series by the first two terms only then: 
\begin{align*}
	E & \ge f_\pi^2 \int {\rm d}^3 x \; \left( \frac{1}{2} \left(\lambda_1^2 + \lambda_2^2 +
		\lambda_3^2 \right) + \frac{1}{8 \beta^2} \left(\lambda_1^2 + \lambda_2^2 +
		\lambda_3^2 \right)^2 \right) \ge f_\pi^2 \int {\rm d}^3 x \; 2\sqrt{\frac{1}{16\beta^2 } \left(\lambda_1^2 + \lambda_2^2 +
		\lambda_3^2 \right)^3} \\
		& \ge  \frac{ f_\pi^2}{2\beta}  \int {\rm d}^3 x \left( 3 \sqrt[3]{\lambda_1^2\lambda_2^2\lambda_3^2}\right)^{3/2} \ge f_\pi^2 \frac{3^{3/2}}{2}  2\pi^2 |B|.
\end{align*}
where the inequality of arithmetic and
geometric means (AM--GM) has been used twice.
\\ 
If we take into account also the third term in the expansion then the bound can be made even tighter. Each term of the sum of the integrand can be bounded using the AM--GM inequality: $\lambda_1^2 + \lambda_2^2 + \lambda_3^2 \ge 
3 | \lambda_1 \lambda_2 \lambda^3 |^{2/3}$. We then want to use the AM--GM in more general form, namely:
\[
	\sum_{i = 1}^n w_i x_i \ge \prod_{i = 1}^n x_i^{w_i} \;,
\]
where $x_i > 0$, $w_i > 0$ and $\sum_{i=1}^n w_i = 1$.\\
This allows us to bound the integrand in the following manner:
\begin{align*}
	\alpha \left(\frac{1}{\alpha} \frac{3}{2} |\lambda_1 \lambda_2 \lambda_3|^{2/3} \right)
	+ \left(\frac{3}{2} - 2 \alpha \right) \left( \frac{1}{\frac{3}{2} - 2 \alpha} \cdot
		\frac{9}{8 \beta^2} |\lambda_1 \lambda_2 \lambda_3|^{4/3}  \right)
	+ \left(\alpha - \frac{1}{2} \right) \left( \frac{1}{\alpha - \frac{1}{2}} \cdot 
		\frac{27}{16 \beta^4} |\lambda_1 \lambda_2 \lambda_3|^{2} \right) \\
	\ge \frac{1}{\beta} \left( \frac{3}{2\alpha} \right)^\alpha 
		\left( \frac{9}{4 (3 - 4\alpha)} \right)^{\frac{3}{2} - 2\alpha}
		\left( \frac{27}{8 (2 \alpha - 1)} \right)^{\alpha - \frac{1}{2}}
		 |\lambda_1 \lambda_2 \lambda_3|\;,
\end{align*}
where $\alpha \in \left[\tfrac{1}{2}, \tfrac{3}{4}\right]$.
Numerical investigation shows that the expression in front of $|\lambda_1 \lambda_2 \lambda_3|$
has its maximum at $\alpha \approx 0.64286$. This finally gives us:
\begin{align*}
	E &\ge f_\pi^2 \cdot \frac{C_3}{\beta} \cdot 2\pi^2 \int {\rm d}^3 x \; \frac{1}{2\pi^2} |\lambda_1 \lambda_2 \lambda_3| \\
	& \ge f_\pi^2 \cdot \frac{C_3}{\beta} \cdot 2\pi^2 |B|.
\end{align*}
where $C_3 \approx 3.5$, compared with previously obtained $C_2\equiv 3^{3/2}/2 \approx 2.6$. 
Taking $\beta = 1$ and $B=1$ we compare our result with the one obtained by Pavlovskii~\cite{pavl1}:
\[
	\frac{E(\beta = 1, B = 1)}{f_\pi^2} = 8\pi \cdot 3.487 \approx 87.638 \ge 69.087 \approx 3.5 \cdot 2\pi^2 \;.
\]
This gives the relative error of our band to be about $21 \%$. Obviously, we may improve the bound by inclusion of next terms in the expansion.  
\section*{Acknowledgemnts}
The author would like to express his special thanks of gratitude to 
A. Wereszczy\'{n}ski for many helpful comments and remarks.
Secondly the author is thankful to C. Adam for derivation of the BPS
equation presented in section II.A.


\end{document}